\documentclass[prd,preprint,a4,superscriptaddress,altaffilletter,showpacs]{revtex4}

\usepackage{graphicx}
\usepackage{bm}
\newcommand{\be}{\begin{equation}}
\newcommand{\ee}{\end{equation}}
\newcommand{\ben}{\begin{eqnarray}}
\newcommand{\een}{\end{eqnarray}}
\newcommand{\n}{\label}
\newcommand{\no}{\noindent}
\newcommand{\la}{\lambda}

\newcommand{\ga}{\gamma}

\begin{document}

\title{Anisotropic k-essence cosmologies}

\author{Luis P. Chimento}
\email{chimento@df.uba.ar}
\author{M\'onica Forte}
\email{monicaforte@fibertel.com.ar}
\affiliation{Departamento de F\'\i sica, Facultad
de Ciencias Exactas y Naturales, Universidad de Buenos Aires,
Ciudad Universitaria, Pabell\'on I, 1428 Buenos Aires,Argentina}

\bibliographystyle{plain}


\begin{abstract} 

We investigate a Bianchi type-I (BTI) cosmology with k-essence and find the set of models which dissipate the initial anisotropy. There are cosmological models with extended tachyon fields and k-essence having constant barotropic index. We obtain the conditions leading to a regular bounce of the average geometry and the residual anisotropy on the bounce. For constant potential, we develop purely kinetic k-essence models which are dust dominated in their early stages, dissipate the initial anisotropy and end in a stable de Sitter accelerated expansion scenario. We show that linear k field and polynomial kinetic function models evolve asymptotically to  Friedmann-Robertson-Walker (FRW) cosmologies. The linear case is compatible with an asymptotic potential interpolating between $V_l\propto \phi^{-\gamma_l}$, in the shear dominated regime, and $V_l\propto\phi^{-2}$ at late time. In the polynomial case, the general solution contains cosmological models with an oscillatory average geometry. For linear k-essence, we find the general solution in the BTI cosmology when the k field is driven by an inverse square potential. This model shares the same geometry as a quintessence field driven by an exponential potential.   

\end{abstract} 


\maketitle

\section{Introduction}  
A number of recent astrophysical data including supernovae Ia \cite{Riess:2004nr}, 
Cosmic Microwave Background \cite{Spergel:2003cb} and large-scale structure 
\cite{Tegmark:2003ud} suggest that we are living in an accelerated universe 
containing about $70 \%$ dark energy, $25 \%$ dark matter and $5 \%$ in other 
forms such as radiation and usual baryonic matter. 
 Among the distinct candidates for dark matter there are: moduli fields 
\cite{Brustein:1998pn}, Wimpzillas \cite{Chung:1998ua}, axinos and 
gravitinos (super partners of the axion and graviton respectively) 
\cite{super}, neutralinos and axions \cite{A7}.  
Simple examples of dark energy  are the cosmological constant and an important 
class of models related with quintessence \cite{quintessence}.
Some of them have the interesting property that the scalar field density remains 
close to the dominant background matter density during most of cosmological 
evolution (tracker fields) \cite{U}. These models, with an adequate potential, 
are able to describe both dark energy and dark matter within a tracker 
framework \cite{tracker}.

    There are several suggested dark energy candidates in the literature, such as vacuum polarization \cite{vacuum}, vector models \cite{Armendariz-Picon:2004pm}, tachyons \cite{tachyon}, 
Chaplygin gas \cite{chaplygin}, k-essence \cite{k}, 
Cardassian expansion \cite{Freese}, quasi-steady state cosmology \cite{Narlikar:2002bk} and scalar-tensor models \cite{tensor}. The k-essence cosmology has received a lot of effort  \cite{Kim}, \cite{Chimento:2003zf},  \cite{Malq}
and recently, it has been shown that it may play both the dark matter and 
dark energy roles \cite{Laforch}. In this way, the k-essence would  explain 
the coincidence problem, namely: {\em Why
are the matter and dark energy densities from the same
order today?}, because  the transition between the tracker behaviour during the radiation-matter domination and a 
cosmological-constant-like behaviour, seems to occur for purely 
dynamical reasons without fine-tuning necessity. 

The results mentioned previously have been established for times after the era in which the universe became 
transparent to radiation and their
    extrapolation to earlier times is totally ungrounded.  There are theoretical
    arguments that support the existence of an anisotropic phase that approaches
    an isotropic one \cite{Chim}. Besides,  if we intend to avoid the 
assumption of special initial conditions tacitly implied in the FRW 
cosmologies,  we should study more
    appropriate models in which anisotropies, perhaps damped out in the course of evolution
 \cite{Hu:1978zd}, can exist from the very beginning. 
With those things in mind we will study  BTI universe filled with k-essence.
In this direction,
we focus on the dissipation of
the initial anisotropy for a wide class of k-essence models which 
asymptotically behave like FRW ones. 
A final contracting singularity is expected for a universe filled with standard matter obeying the energy conditions.
Dissipative effects of particle creation could violate the energy conditions avoiding 
the final singularity with a regular bounce. 
So, it will be interesting to investigate this subject when there is a k-essence 
source in an anisotropic background. 

Finally, we compare different cosmological models like that associated with quintessence or k-essence in a  BTI space time,
contributing to gain more insight
on the degree of resemblance between quintessence
and k-essence scenarios driven by exponential and inverse square  potentials respectively \cite{q=k}. In section II, we state the Einstein equations in  BTI space time with k-essence. In section III, we study the dissipation of the anisotropy, the existence of an average bounce and
find the first integral of the k-field equation for  purely kinetic k-essence models. In section IV, we obtain the general solution of the Einstein equations in two cases: a ) when the k field depends linearly on the cosmological time and b ) when the barotropic index of the k-essence is constant. In section V, we find the general solution of the Einstein equation for a linear k-essence driven by an inverse square potential and show the geometrical equivalence of this model with the  quintessence one driven by an exponential potential. Finally, in section VI we present our conclusions.

\section{ BTI k-essence cosmologies}

In the  BTI space time described by the line element  
\be
\n{m}
ds^2=-dt^2+a_1^2(t)dx_1^2+a_2^2(t)dx_2^2+a_3^2(t)dx_3^2,  
\ee  

\no with a perfect fluid having energy density $\rho$ and isotropic pressure $p$, the Einstein equations are given by
\ben
\n{2}
3H^2 = \rho + \frac{1}{2}{\sigma^2}\\
\n{3}
-2 \dot H = p+\rho+ \sigma^2\\
\n{dc}
\dot\rho+3H(\rho+p) = 0\\
\n{dd}
\dot{\sigma} + 3H{\sigma}=0
\een

\no where $H$ is the average of the individual expansion rates $H_1={\dot a_1}/{a_1}, H_2={\dot a_2}/{a_2}, H_3={\dot a_3}/{a_3}$. They are related to the three spatial directions and 
\be
\n{H]}  
H = \frac{\dot a}{a} = \frac{1}{3} (H_1 + H_2 + H_3), 
\ee

\no while the scale factor $a$ is defined as
\be
\n{e}
a = (-g)^{1/6} = (a_1a_2a_3)^{1/3}.
\ee

\no Here and throughout overdots denote differentiation with
respect to the cosmological time $t$ and we assume that $8\pi G=1$. The shear vector $\vec{\sigma}$ has components 
\be
\n{si}
\sigma_i = H_i - H,
\ee

\no where
\be
\n{s}
\sigma_i=\frac{\sigma_{i0}}{a^3},  
\qquad \sigma=\sqrt{{\vec\sigma}\cdot{\vec\sigma}}=\frac{\sigma_0}{a^3},
\ee

\no and the three constants $\sigma_{i0}$ transform as the
components of a vector in the internal three-dimensional cartesian space associated with the three axis $\sigma_i$. They satisfy the transverse
condition 
\be
\n{s1}
\sigma_{10}+\sigma_{20}+\sigma_{30}=0,
\ee
with
\be
\n{s2}
\sigma^2_{10}+\sigma^2_{20}+\sigma^2_{30}=\sigma_0^2,
\ee

\no and $\sigma_0$ a constant. Inserting the first Eq. (\ref{s}) into
 Eq. (\ref{si}) and integrating, we obtain the individual scale factors $a_i$
\be
\n{ai}
a_i=\,a_{i0}\left(\frac{3\sigma_0^2}{2}\right)^{1/6}\,\frac{m^{s_i}}{\dot m^{1/3}}, 
\ee

\no where the integration constants $a_{i0}$ satisfy the condition $a_{10}a_{20}a_{30}=1$, the function $m$ is defined through 
the following equation
\be
\n{m}
\frac{\dot m}{m}=\sqrt{\frac{3}{2}}\,\,\frac{\sigma_0}{a^3},
\ee

\no and the parameters $s_i$ satisfy the Kasner constraints \cite{BI}
\be
\n{ps}
s_1+s_2+s_3=1, \qquad s^2_1+s^2_2+s^2_3=1.
\ee

\no A one-parameter representation of the Kasner exponents $s_i$ is given by
$$ 
s_1=\frac{1}{3}\left[1+\frac{2u}{\sqrt{3+u^2}}\right], \,
s_2=\frac{1}{3}\left[1+\frac{3-u}{\sqrt{3+u^2}}\right],
$$ 
\be
\n{pb}
s_3=\frac{1}{3}\left[1-\frac{3+u}{\sqrt{3+u^2}}\right].
\ee

\no where $u$ is the parameter. 

Let us assume that the anisotropic space time contains an isotropic  perfect
 fluid associated with the k-essence field $\phi$.
  The latter is  introduced by means of a Lagrangian factorized in the following way
\be
\n{l}
\it{L}=-V(\phi)F(x),
\ee

\no where the potential $V ( \phi )$ is a positive definite function
depending on the k field ${\phi}$ and $F$ is a
function of the kinetic term $x = g^{ik}\phi_i\phi_k$, with
$\phi_i=\partial\phi/\partial x^i$. Identifying the energy-momentum tensor of the k field with that of a perfect fluid, the energy density ${\rho_\phi}\ $ and the pressure $p_{\phi}$ are 
\be
\n{UF}
\rho_\phi = V(F - 2xF_x), \qquad p_{\phi}= -VF,
\ee

\no with $F_x=dF/dx$. Assuming the equation of state $p_{\phi}=
(\ga_\phi - 1)\rho_\phi$, the barotropic index $\ga_\phi$
\be
\n{BI}
\ga_\phi=\frac{-2xF_x}{F-2xF_x},
\ee

\no depends only on the kinetic term.

The dynamic of the k field in the  BTI space time is obtained inserting the energy density and pressure (\ref{UF}) into Eqs. (\ref{2})-(\ref{dd}) and (\ref{BI}). They become
\ben
\n{da}
3H^2 = V(F - 2xF_x) + \frac{\sigma_0^2}{2a^6},\\
\n{db}
-2 \dot H = -2VxF_x +\frac{\sigma_0^2}{a^6},\\
\n{kgk}
[F_x+2xF_{xx}]\ddot\phi+3HF_x\dot\phi+\frac{V'}{2V}[F-2xF_x]=0,\\
\n{gag}
\gamma_\phi=-\frac{2\dot H+\sigma_0^2/a^6}{3H^2-\sigma_0^2/2a^6}.
\een

\no Eqs. (\ref{kgk}) and (\ref{gag}) show that $\phi$ and $\ga_\phi$ are sensitive to the evolution of the average geometry.

\section{General issues}

In this section we are going to investigate several interesting problems 
concerning the anisotropic background. These are related to the dissipation 
of the anisotropy, the avoidance of the special initial conditions, the existence 
of a regular bounce and the possibility of developing a kinetic k-essence 
cosmology. All of them share in common a description in terms of the barotropic index independent of the  potential we choose.

\subsection{Dissipation of the Anisotropy}

There is observational evidence that at present the universe seems homogeneous and has been highly isotropic after the recombination era. Observations of the cosmic microwave background radiation reveal that our universe is remarkably uniform, at very large scale, and is currently under accelerated expansion. Dissipation of anisotropy could solve the
problem of the apparent large-scale observed isotropy of the universe.
A realistic model should describe the existence of an initial anisotropic phase that approaches an isotropic one for any initial condition. It suggests to study cosmological models in which anisotropies, existing at early stage of the expansion, are damped out in the course of the evolution. This investigation has increased since it was shown in Ref. \cite{Hu:1978zd} that the creation of scalar particles can dissipate the anisotropy as the universe expands. The dissipation of the anisotropy in a  BTI universe is particularly interesting because it is the simplest model that includes the isotropic FRW universe. 
Here, we investigate this process when the universe is filled with k-essence. 
It brings the possibility of considering dynamical or purely kinematic models. 


To investigate the dissipation of the initial anisotropy by the expansion of the universe, we introduce a positive magnitude
defined by the ratio of the energy density contribution of the shear $\sigma^2/2$ and the k-field energy density $\rho_\phi$. Hence,  differentiating $D=\sigma^2/2\rho_\phi$ with respect the cosmological time $t$ and using Eqs. (\ref{dc})-(\ref{dd}) and (\ref{BI}), we obtain the evolution equation for the ratio $D$
\be
\n{dissi}
\dot D+3H(2-\ga_\phi)D=0.
\ee

\noindent For an average expanding cosmology ($H>0$) the solution of the last equation,  $D=0$, is asymptotically stable whenever $D$ is a positive definite quantity and $\ga_\phi<2$. In this asymptotic regime the fluid dominates compared with the shear. These models, generated by kinetic functions satisfying the condition 
\be
\n{c1}
\frac{F-xF_x}{F-2xF_x}>0,
\ee

\no become isotropic at late times and the geometry tends to the typical FRW one, whatever the potential be. On the other side, using the condition   $\ga_\phi<2$ in Eqs. (\ref{BI})-(\ref{db}) we get $\dot H+3H^2>0$, which combined with Eqs. (\ref{2}) and (\ref{3}), finally gives $\rho>p$.
So, fluids obeying the dominant energy condition (DEC) dissipate the initial anisotropy without using a selected initial condition. In contrast, when $\ga_\phi>2$ the shear dominates compared with the fluid, the DEC is violated and the magnitude $D$ increases asymptotically.

For instance, the extended tachyon fields generated by \cite{Chim}
\be
\n{fo}
F_r^\mp=\left[1\mp(-x)^r\right]^{1/2r},   
\ee

\no have sound speed $c_{s}^2=(1-\ga_\phi)/(2r-1)$. They generalize the standard tachyon field yielded by  $F_1^-$. $F_r^+$ represents k-essence with negative pressure and negative barotropic index which can be identified with phantom matter. For large $r$, the fluids associated with $F_r^-$ satisfy the equation of state $p=-\rho=-V$ and behave like a variable cosmological constant depending on the k field. The set of extended tachyon fields $(\ref{fo})$ verify the condition (\ref{c1}) and dissipate the initial anisotropy along the evolution of the universe. 
	
Another interesting case is provided by the polynomial kinetic function
\cite{Chimento:2003zf}
\be
\n{Fga}
F_{\gamma_p}^\pm (x)=\pm (-x)^{\ga_p/2(\ga_p - 1)}, \qquad \ga_p\ne 1, 
\ee

\no
which yields a constant barotropic index $\ga_\phi=\ga_p$ and an energy density $\rho_\phi\propto 1/a^{3\ga_p}$, generalizing the usual perfect fluids. The set of fluids with $\ga_p<2$ fulfill the condition (\ref{c1}) and the cosmological models generated by $F_{\ga_p}^\pm$ have a final isotropic stage. 
 
\subsection{Average bounce}  

Let us consider the problem of singularities arising in a final contracting universe. For a universe filled with standard matter obeying the energy conditions, a singularity is expected as a consequence of that contraction. Quantum
effects, through dissipative effects of particle creation, could violate the
energy conditions, raising the
possibility of avoiding the final singularity with a regular bounce. The existence of singularity-free contracting solutions is important to describe an eternal oscillating model of the universe. Different examples of singularity-free cosmological models with a regular bounce have been considered in Refs. \cite{Ma}-\cite{MaMu}. It seems reasonable to gain insight in this problem by considering other kind of frameworks. In particular, we focus on cosmological models with a k-essence source.

For a BTI space time an average bounce will occur at the time $t=t_0$,
 where the average expansion rate is $H(t_0)=0$, the scale factor is stationary $\dot a(t_0)=0$ and $a(t_0)\ne 0$ \cite{Frolov:1991}. Then, from Eq. (\ref{da}) the energy density of the k field is negative during a finite time interval around $t_0$ and satisfies, at $t_0$, the condition
\be
\n{rb}
V(F-2xF_x)=-\frac{\sigma^2}{2}.
\ee

\no Also, we assume that the scale factor has a minimum at the average bounce,
$\ddot a ( t_0 ) >0$, so $\dot H ( t_0 ) >0$. Then from Eqs. (\ref{db}) and (\ref{rb}), we get
\be
\n{pb}
VF>\frac{\sigma^2}{2}.
\ee

\no The restrictions on the energy density and the pressure of the k field, expressed by Eqs. (\ref{rb}) and (\ref{pb}), give the lower limit $\ga_\phi>2$ and 
\ben
\n{fb}
F>0,  \qquad F_x<0,\\
\n{fxb}
xF_x<F<2xF_x.
\een

\no Conditions (\ref{fb}) and (\ref{fxb}) are fulfilled by the branch $F_{\ga_p}^+$ in (\ref{Fga}) with $\ga_p>2$. In the next section we will find the general solution of Eqs. (\ref{da})-(\ref{kgk}) for the two branches (\ref{Fga}) and show that the $F_{\ga_p}^+$ branch generates an oscillatory average scale factor. 

The existence of an average bounce, defined by the conditions $H=0$ 
and $\dot H>0$, is linked to atypical perfect fluids with $\ga_\phi>2$. This problem can be overcame in the k-essence framework. However, on the average bounce the relation between the shear and the k field   energy densities is constant $D=-1$. Hence, we can reach a bounce avoiding the final contracting singularity but we have a residual anisotropy in that scenario and $D$ becomes  unstable.

In summary, there is a strict incompatibility between the stability 
condition and the existence of an average bounce. In fact, the former occurs when the fluid that constitutes the source satisfies the DEC and the latter happens when it is violated. 
That incompatibility is due to the requirement D>0. (that is, positive energy 
    density), in order to apply the Lyapunov  theorem in Eq. (\ref{dissi}). 
In fact, an average bounce exists if the energy density and $D$ become 
negative, see Eq. (\ref{rb}).

\subsection{Kinetic k-essence cosmology}
               
Here, we consider a constant potential $V=V_0$ and investigate the resulting purely kinetic k-essence cosmologies in the BTI background. Models of this class are: a) the generalized Chaplygin gas derived from a Lagrangian containing non-standard kinetic-energy terms and proposed as unified dark matter, b) the modified and extended Chaplygin gases which were suggested as alternatives to the above model \cite{Chim}. For a constant potential, the k field Eq. (\ref{kgk}) can be rewritten as
\be
\n{vcte}
\left(\frac{\gamma_\phi}{\dot\phi}\right)^.+
3H\left(\frac{\gamma_\phi}{\dot\phi}\right)(1-\gamma_\phi)=0.
\ee
 
\no Using the geometrical definition (\ref{gag}) of $\ga_\phi$ in Eq. (\ref{vcte}), we get its general first integral 
\be
\n{pi1}
\left(\frac{\gamma_\phi}{\dot \phi}\right) =  \frac{ c}{a^3(3H^2 - \sigma_0^2/2 a^6)},
\ee
for any function $F$. Using again the expression for $\gamma_\phi$ in the last equation, it becomes
\be 
\n{dotphi}
\dot\phi=-\frac{a^3}{c}(2\dot H+\frac{\sigma_0^2}{a^6}) = \frac{a^3}{c}\gamma_\phi\rho_\phi,
\ee
or 
\be
\n{igFRW}
a^3 \dot\phi F_x = \frac{c}{2V_0},
\ee

\no after replacing $\gamma_\phi\rho_\phi=2V_0{\dot\phi}^2 F_x$.

From Eqs. (\ref{BI}), (\ref{dotphi}) and (\ref{igFRW}), the barotropic index
associated with this purely kinetic k-essence becomes
\be
\n{gam}
\gamma_\phi=(1+\frac{2V_0^2\sigma_0^2FF_x}{c^2\sigma^2})^{-1}.
\ee
The models generated by the set of kinetic functions satisfying the condition $FF_x/\sigma^2\ll 1$ at early times, describe universes which on the average are dust dominated in their early stages, that is, $\gamma_\phi\approx 1$, $p_\phi\approx 0$ and $\rho_\phi\approx
a^{-3}$. Such models dissipate the initial anisotropy and the universe ends in a stable de Sitter accelerated expansion scenario. These transient models resemble the generalized, modified and extended Chaplygin gases, i.e., they interpolate between dark matter at early times and dark energy at late times. 

\section{Asymptotic power law solutions}

In FRW cosmologies there are two different ways of obtaining power law solutions. In the first approach, the k field depends linearly on the time $\phi=\phi_0 t$, and $x=x_0=
 -{\dot\phi}^2=-\phi_0^2$ is a constant. This means that $F(x_0)=f$ and $F_x(x_0)=f'$ are constants for any $F$, while the potential becomes 
\be 
\n{v2}
V=\frac{V_0}{\phi^2}.
\ee
 
\no In the second approach, one imposes that $\ga_\phi=\ga_p$ is a constant, so Eq. (\ref{BI}) becomes a differential equation for $F$ and their solutions are given by the polynomial functions (\ref{Fga}). Also, the conservation equation can be integrated giving a constraint
between the potential $V_{p}$, $F^\pm_{\ga_p}$ and the scale factor
\be
\n{vf}
\rho_{\phi_p}=\frac{V_{\ga_p}F^\pm_{\ga_p}}{1-\ga_p}=\frac{\la_p}{a^{3\gamma_p}},
\ee

\no where $\la_p$ is an integration constant.

\subsection{The anisotropic background}

Now, we will perform the above analysis investigating linear k-field and polynomial kinetic function cases in the anisotropic  BTI background and the results will be compared with that of the isotropic FRW background. 

In the linear case, the Einstein Eqs. (\ref{da})-(\ref{kgk}) read 
\ben
\n{da'}
3H^2 = \frac{V_lf}{1-\ga_l} + \frac{\sigma_0^2}{2a^6}\\
\n{db'}
-2 \dot H = 2V_l\phi_0^2 f' +\frac{\sigma_0^2}{a^6}\\
\n{kg'}
3\ga_l H+\frac{V_l'}{V_l}\phi_0=0,
\een 

\no where we have used the barotropic index (\ref{BI}) evaluated on the linear k field
\be
\n{gl}
\ga_l=\frac{2\phi_0^2 f}{f+2\phi_0^2 f'}.
\ee

\no Assuming that the scale factor is a function of $\phi$, then $H=\phi_0 da/ad\phi$ and the conservation Eq. (\ref{kg'}) can be integrated, obtaining 
\be
\n{av}
\rho_{\phi_l}=\frac{V_lf}{1-\ga_l}=\frac{\la_l}{a^{3\ga_l}},
\ee

\no where $\la_l$ is an integration constant. Back to Eq.(\ref{da'}), it can be rewritten as 
\be
\n{00l}  
3H^2=\frac{\la_l}{a^{3\ga_l}}+\frac{\sigma_0^2}{2a^6}.
\ee

When the k field is generated by the set of polynomial functions (\ref{Fga}), the dynamics of the scale factor $a$ is obtained by inserting the energy density (\ref{vf}) into the Eq. (\ref{da}), so
\be
\n{00p}
3H^2=\frac{\lambda_p}{a^{3\ga_p}}+\frac{\sigma_0^2}{2a^6}.
\ee

\no Eqs. (\ref{00l}) and (\ref{00p}) show that both cases reduce to a BTI
cosmology with a perfect fluid having energy density and barotropic index 
not necessarily positive definite. 

\subsection{Metric and potentials}

Introducing the new variable $v=a^3=\sqrt{-g}$ into Eqs. (\ref{00l}) and (\ref{00p}), they become
\be
\n{lof}
v^{'2} = 1 + \frac{2\la}{\sigma_0^2}v^{2 - \gamma}
\ee

\no where the prime indicates differentiation with respect to the dimensionless time $\ T=\sqrt{3/2}\sigma_0 t$ and $\la,\ga=\la_l,\ga_l$ or $\la_p,\ga_p$ respectively. The general solution of Eq. (\ref{lof}) depends on the sign of the integration constant $\la$, being hyperbolic for $\la>0$ and oscillatory for $\la<0$. There are three kinds of solutions:

\begin{enumerate}
\item
$\ga=2$
\be  
\n{ni}
a^3= \sqrt{3\left(\la+\frac{\sigma_0^2}{2}\right)}\,\, t,
\ee
\be
\n{nii}
a_i=a_{0i} \left[3(\la + \frac{\sigma_0^2}{2})\right]^{1/6} 
t^{1/3+(s_i-1/3)/\sqrt{1+2\la/\sigma_0^2}}
\ee

\item
$\la>0$
\be
\n{nme}
a^{3(2-\ga)}= \frac{\sigma_0^2}{2\la}\sinh^2\tau,  
\quad t = t_0^+\int (\sinh\tau)^{\ga/(2-\ga)} d\tau,  
\ee
\be
\n{nmei}
a_i=a_{0i}\left(\frac{2\sigma_0^2}{\la}\right)^{1/3(2-\ga)}\left[\cosh\frac{\tau}{2}\right]^{4/3(2-\ga)}\left[\tanh\frac{\tau}{2}\right]^{2s_i/(2-\ga)}
\ee

\item 
$\la< 0$
\be  
\n{nma}
a^{3(2-\ga)}=\frac{\sigma_0^2}{-2\la}\sin^2\tau, 
\quad t = t_0^-\int (\sin\tau)^{\ga/(2-\ga)} d\tau,  
\ee
\be  
\n{nmai}
a_i=a_{0i}\left(\frac{2\sigma_0^2}{-\la}\right)^{1/3(2-\ga)}\left[\cos\frac{\tau}{2}\right]^{4/3(2-\ga)}\left[\tan\frac{\tau}{2}\right]^{2s_i/(2-\ga)}
\ee 
\end{enumerate}

\no where $t_0^\pm=2^{3/2}(\pm\sigma_0^2/2\la)^{1/(2-\ga)}[\sqrt{3}
\sigma_0(2-\ga)]^{-1}$.

>From Eq. (\ref{00l}), the linear k field can be expressed as a function of the scale factor
\be
\n{Vlin}
\phi=\frac{\sqrt{6}\,\, \phi_0}{\sigma_0} \int \frac{a^2 da}{\sqrt{1 + 2 \lambda_l a^{3(2 - \gamma_l)}/\sigma_0^2}}.
\ee

\no Also, for  $0<\gamma_l<2$, the Eq. (\ref{nme}) shows that $t$ and $\tau$ have the same asymptotic limits. In this case, the last equation is appropriate to investigate the relation between $\phi$ and $V_{l}$ in the two asymptotic regimes. In the first regime, $a^{3(2-\ga_l)}<\sigma_0^2/2\lambda$, the shear dominates compared with the perfect fluid and $\phi\propto a^3$. Hence, (see Eq. (\ref{av})) the potential becomes $V_l\propto \phi^{-\gamma_l}$ and qualitatively, the universe has a Kasner scenario. In the second regime, which starts
from some characteristic time where $a^{3(2-\ga_l)}>\sigma_0^2/2\lambda$ the fluid becomes dominant and $\phi\propto a^{3\gamma_l/2}$. It leads to the asymptotic inverse square potential $V_l\propto\phi^{-2}$. Hence, owing to the spatial isotropy of the stress-energy tensor, the anisotropic  BTI model evolves into a FRW cosmology and the initial anisotropy of this model is dissipated as the universe expands.

In the polynomial function case, the scale factor tends asymptotically to a power law solution (for $\ga_p=2$, stiff fluid, we get the scale factor (\ref{ni})). In contrast, the same model in the FRW space time leads to exact power law solutions. Using Eqs. (\ref{Fga}), (\ref{vf}), (\ref{ni}),(\ref{nme}) and (\ref{nma}), we find the following relations between $V_{\ga_p}$ and the time $\tau$
$$
\int d\phi V^{(\ga_p-1)/\ga_p}_{\ga_p}= 
$$
\ben
\n{campo}
\sqrt{\frac{2}{3}}
\frac{\sigma_o}{\la_p(2-\ga_p)} {|\la_p(1-\ga_p)|}^{(\ga_p-1)/\ga_p}\zeta(\tau),  \quad \ga_p\ne 2,\\
\n{g=2}
\int d\phi\sqrt{V_{\ga_p}}=\frac{1}{\sqrt{|-3(1+\sigma_0^2/2\la_p)|}}\ln{t},
\quad  \ga_p=2,
\een

\noindent where $\zeta(\tau)=\cosh{\tau}$ for $\lambda_p>0$ or $\zeta(\tau)=\cos{\tau}$ for $\lambda_p<0$. For purely kinetic k-essence models, $\phi$ becomes proportional to $\zeta(\tau)$, or $\ln{t}$ according to $\ga_p\ne 2$ or $\ga_p=2$. 

\section{The inverse square potential}

In \cite{Chim} it was shown that k-essence driven by an inverse square potential in FRW space time and generated by $F=1+mx$, with $m$ a constant, is kinematically equivalent to quintessence  driven by an exponential potential. In this section, we will extend this conclusion to the anisotropic  BTI cosmology. To do that, we use Eqs.  (\ref{UF}) and (\ref{BI}) to rewrite the k field Eq. (\ref{kgk}) as 
\be
\n{ga.}
\left(\frac{\gamma_\phi}{\dot\phi}\right)^.+
3H\left(\frac{\gamma_\phi}{\dot\phi}\right)(1-\gamma_\phi)+\frac{V'}{V}(1-\gamma_\phi)=0.
\ee

\no where $V'=dV/d\phi$. Substituting in the last equation
\be
\n{pi}
\gamma_\phi=-\frac{\dot V}{V}\left(\frac{H}{\rho_\phi}+L\right),
\ee

\no where $L$ is an arbitrary function and using Eqs. (\ref{BI})-(\ref{db}), it becomes
\be
\n{l.}
\dot
L+3HL(1-\gamma_\phi)+\left[\frac{1}{2}+
\left(\frac{V}{V'}\right)^{'}\,\right]\gamma_\phi=0.
\ee

\no Choosing the potential (\ref{v2}), the latter reduces to a differential equation containing only geometrical quantities. Its general solution is
\be
\n{lg}
L=-\frac{3c}{2a^3}\left(3H^2-\frac{\sigma_0^2}{2a^6}\right)^{-1},
\ee

\no where $c$ is an integration constant and we have used the
geometrical definition (\ref{gag}) of $\ga_\phi$. Finally, the first integral of the k field  Eq. (\ref{kgk}) or (\ref{ga.}) can be written in three different ways as
\ben
\n{PI1}
\left(\frac{\gamma_\phi}{\dot \phi}\right) \phi = \frac{H^2}{(3H^2 - \sigma_0^2/2 a^6)}\left(\frac{2}{H} + \frac{3 c}{a^3 H^2}\right),\\
\n{PI2}
\dot \phi F_x - ( H + \frac{3 c}{2 a^3})\frac{\phi}{V_0}=0,\\
\n{PI3}
- V_0 F_x (\dot H + \frac{{\sigma_0}^2}{2 a^6}) = ( H + \frac{3 c}{2 a^3})^2,
\een

\no for any function $F$. 

Inserting $F=1+mx$ in the Eq. (\ref{PI3}), we obtain the following equation for the scale factor $a$
\be
\n{SFK}
s'' + s^{\alpha}s' + \left( \frac{1}{4} + \frac{mV_0{\sigma}_0^2}{18 c^2}\right) s^{2\alpha +1} = 0,
\ee

\no
where we have used the new variables $s$ and $\xi$ defined by
\be
\n{st}
s=a^{-3/\alpha},  \qquad \xi=\frac{3c\,t}{mV_0},
\ee

\no with $\alpha= - 3 m V_0$. Eq. (\ref{SFK}) is a particular case of a more general equation investigated and solved in \cite{f}. Following this reference, we introduce the change of variables
\ben
\n{zs}
z=\frac{s^{\alpha+1}}{\alpha+1}, \qquad \alpha \ne -1,\\
\n{z-1}
z=\ln{s}, \qquad \alpha=-1,\\
\n{e}
\eta=\int s^{\alpha}d\xi,
\een

\no in Eq. (\ref{SFK}). Then, it transforms into two second order linear differential equations
\ben
\n{desig}
\frac{d^2 z}{d \eta^2} + \frac{d z}{d \eta} + (\alpha + 1)\left(\frac{1}{4} -  \frac{\alpha \sigma_0^2}{54 c^2}\right) z  = 0,  \quad \alpha \ne -1,\\
\n{igual}
\frac{d^2 z}{d \eta^2} + \frac{d z}{d \eta} + \left(\frac{1}{4} + \frac{\sigma_0^2}{54 c^2}\right) = 0, \quad \alpha = -1,
\een
 
\no whose explicit solutions are: 

\begin{enumerate}
\item
\no $\alpha>0$
\ben
\n{sol1}
a(\eta)= \left[\sqrt{B} \,\, e^{-\eta/2}\sin(\sqrt{\mu}\,\eta + \eta_0)\right]^{-\alpha/3(\alpha + 1)},\\
\n{s1i}
a_i= a_{0i}\,\frac{e^{-(\beta+\sigma_{_0}\alpha s_i/\sqrt{6}\,c)\eta/3}}
{\left[ \sqrt{B}  \sin(\sqrt{\mu}\,\eta + \eta_0)\right]^{\alpha/3(\alpha + 1)}}.
\een

\item
$-1<\alpha<0$
\ben
\n{sol2}
a(\eta)= \left[  \sqrt{B} e^{-\eta/2}\cosh(\sqrt{-\mu}\,\eta + \eta_0)\right]^{-\alpha/3(\alpha + 1)},\\
\n{s2i}
a_i=  a_{0i}\,\frac{e^{-(\beta+\sigma_{_0}\alpha s_i/\sqrt{6}\,c)\eta/3}}
{\left[ \sqrt{B}\cosh(\sqrt{-\mu}\,\eta + \eta_0)\right]^{\alpha/3(\alpha + 1)}}.
\een

\item
$\alpha=-1$
\ben
\n{sol3}
a(t)=a_0e^{-(1/2+\sigma_0^2/27 c^2)\eta/6+V_0e^{-\eta}/81 \phi_0^2 c^2},\\
\n{s3i}
a_i= a_{0i}a_0e^{-((1/2+\sigma_0/3\sqrt{6}c)^2+\sigma_0 s_i/\sqrt{6}c)\eta/3+V_0e^{-\eta}/81\phi_0^2 c^2}. 
\een

\item
$\alpha<-1$
\ben
\n{sol4}
a(\eta)= \left[  \sqrt{-B} e^{-\eta/2}\sinh(\sqrt{-\mu}\,\eta + \eta_0)\right]^{-\alpha/3(\alpha + 1)},\\
\n{s4i}
a_i=  a_{0i}\,\frac{e^{-(\beta+\sigma_{_0}\alpha s_i/\sqrt{6}\,c)\eta/3}}{\left[ \sqrt{-B}\sinh(\sqrt{-\mu}\,\eta + \eta_0)\right]^{\alpha/3(\alpha + 1)}},
\een

\end{enumerate}

\no with $a_{01}a_{02}a_{03}= 1$ and 
\ben
\n{mu}
\mu=\frac{\alpha}{2}\left[\frac{1}{2} - \frac{\sigma_0^2}{27 c^2}(\alpha + 1)\right],\\
\n{B}
B=\frac{4(\alpha + 1)V_0}{\phi_0^2[27 c^2 - 2(\alpha + 1)\sigma_0^2]}\,,\\
\n{Up}
\beta = - \alpha\left[\frac{1}{2(\alpha+ 1)}+\frac{\sigma_0}{3\sqrt{6}c} \right].
\een
Writing the above solutions, we have used $\phi$ obtained by integrating
the Eq. (\ref{PI2})
\be
\n{klink}
\phi = \phi_0 a^{-3/\alpha}e^{\eta /2}.
\ee

Now, we consider a  BTI cosmological model with a quintessence field $\varphi$ driven by a potential $U(\varphi)$ and having energy density and pressure given by
\be
\n{rpq}
\rho_\varphi=\frac{q}{2}\dot\varphi^2+U(\varphi), \qquad p_\varphi=\frac{q}{2}\dot\varphi^2-U(\varphi),
\ee

\no where $q$ is a constant. For $q<0$ we have a phantom scalar field while for $q>0$ we have a non-phantom one. Now, the Einstein Eqs. (\ref{2})-(\ref{dd}) become
\ben
\n{00}
3H^2=\frac{q}{2}\dot\varphi^2+U(\varphi)+\frac{\sigma_0^2}{2a^6},\\
\n{.H}
-2\dot H=q\dot\varphi^2+\frac{\sigma_0^2}{a^6},\\
\n{kg}
\ddot{\varphi}+3H\dot{\varphi}+\frac{dU}{qd{\varphi}} = 0.
\een

\no For a quintessence field driven by the exponential potential 
\be
\n{ex}
U=U_0e^{-qA\varphi},
\ee

\no where $A$ is a constant, it can be easily seen that
\be 
\n{.v}
\dot{\varphi} = A H+\frac{b}{a^3},
\ee

\no
with $b$ a constant, is a first integral of the Klein-Gordon Eq. (\ref{kg}). In addition, from Eqs. (\ref{.H}) and (\ref{.v}) we obtain the dynamic equation for the scale factor
\be
\n{..a}
-2\dot H=A^2H^2+2bA\frac{H}{a^3}+\frac{b^2+\sigma_0^2}{a^6}.
\ee

\no After introducing the new variables 
\be
\n{st'}
s=a^{-3/\nu}, \qquad \zeta=qbAt, \qquad  \nu=-\frac{6}{q A^2},
\ee

\no we get the final equation
\be 
\n{SFQ}
s'' + s^{\nu}s' + \left( \frac{1}{4} + \frac{{\sigma}_0^2}{4qb^2}\right) s^{2\nu +1} = 0.
\ee

\no Formally, Eqs. (\ref{SFK}) and (\ref{SFQ}) are similar, so, making the following identification between the parameters
\be
\n{Rel}
mV_0=\frac{2}{qA^2},  \qquad  \frac{3c}{2}=\frac{b}{A},
\ee

\no they become the same. This means that both models are described by the same scale factor and they are geometrically equivalent. 

Turning to Eqs. (\ref{2}) and (\ref{3}) we obtain
\be
\n{mif}
3H^2 + \dot H = \frac{\rho - p}{2}
\ee

\no
where $\rho$ and p are the density of energy and pressure of either the k-essence or the quintessence. Hence, the latter becomes
\be
\n{mifs}
3H^2+\dot H =U({\varphi})=V({\phi})
\ee

\no which leads to
\be
\n{uv}
U({\varphi}(t))=U_0\,e^{-q A \varphi}=V({\phi}(t))=\frac{V_0}{\phi ^2},
\ee

\no and 
\be
\n{campos}
\phi = \sqrt{\frac{V_0}{U_0}}\,\,e^{q A \varphi / 2}.
\ee
\no 
Inserting the k field (\ref{klink}) in this relation, we find the scalar field by simple algebra.        


\section{Conclusions}  

We have written the Einstein equations in a  BTI space time filled with k-essence and 
investigated the cosmological models which dissipate  the initial anisotropy or have a 
regular average bounce. When the barotropic index of the k-essence satisfies $\ga_\phi<2$ 
the fluid dominates compared with the shear and the average expanding cosmology tends 
asymptotically to the stable isotropic FRW cosmology without fine-tunnig necessity. 
These fluids include the extended tachyon fields generated 
by the functions (\ref{fo}) and the polynomial kinetic functions (\ref{Fga}) with 
$\ga_p<2$. In both cases the initial anisotropy is dissipated along the evolution of 
the universe and it has a final isotropic stage. 
However, when $\ga_\phi>2$ the shear dominates compared with the fluid, the DEC is 
violated and the asymptotic condition of stability no longer holds. A universe filled 
with this atypical perfect fluid avoids the final singularity with a regular bounce. 
However, on the average
bounce the relation between the shear and the k field
energy densities is constant, this scenario has a residual anisotropy and  becomes 
unstable. Therefore, the dissipation of the anisotropy in the BTI cosmology and the 
existence of an average bounce of the geometry are different cosmological scenarios. The former is associated with fluids satisfying the DEC and the latter   
with fluids that  violate the DEC. For instance, the polynomial functions (\ref{Fga}) 
$F_{\gamma_p}^+$ with $\ga_p>2$ give rise to a regular bounce.

Astrophysical evidences show that the evolution of the universe is
driven by dark energy with negative pressure together with pressureless cold dark matter. 
We have speculated that a single component acted as both dark matter and dark 
energy. In fact, for a constant potential, we have obtained the first integral of the 
k field equation and shown 
a set of purely kinetic k-essence models describing universes, which on the average are 
dust dominated in their early stages. These models with non-canonical standard kinetic 
terms provide a unified description of dark matter and dark energy. They dissipate the 
initial anisotropy and the universe ends in a stable de Sitter accelerated expansion 
scenario. These transient models, interpolating between dark matter at early times and 
dark energy at late times are promising candidates for quartessence and appear as 
alternatives to the two unifying candidates in the literature, the Chaplygin gas and the 
tachyon field.
The unification of those two components makes model building considerably more simple. 


The construction of cosmological models with tracker behaviour
where the k-essence mimics the equation of state of the radiation-matter component, 
represent fluids with a constant barotropic index. These fluids, give rise to an 
asymptotically power-law behaviour of the scale factor when the underlying geometry is BTI.
 We have studied those asymptotic power law solutions for k-essence models in two different
 cases: i) the k field evolves linearly with the cosmological time and ii) the k field is 
generated by polynomial kinetic functions. Both cases reduce to a  BTI cosmology with a 
perfect fluid and share the same average geometry. A large set of these models evolve into 
a FRW cosmology and the initial anisotropy is dissipated as the universe expands. In the 
linear case, for $0<\la_l<2$, the asymptotic potential interpolates between 
$V_l\propto \phi^{-\gamma_l}$,
in the shear dominated regime and $V_l\propto\phi^{-2}$ at late time. 
This allows to enlarge the dynamics leading to a bigger set of cosmological model with tracker behavior 
in comparison with similar models in FRW cosmology. In the polynomial function case the 
scale factor tends asymptotically to a power law solution in contrast with FRW space times,
 where the same kind of model leads to an exact power law solution. Besides, we have 
shown that the $F_{\ga_p}^+$ branch generates an oscillatory average geometry. 

 
Accelerated expansion is a necessary condition to resolve many basic issues in the present 
cosmology. In general, there is a lot of interest in those models which lead to 
accelerated power-law solutions in FRW cosmology. Such expansion is normally driven by a scalar field
with the Liouville form (exponential potential). 
In a previous paper \cite{Chim}, one of 
us investigated the connections between the inverse square potential and the exponential 
potential when the universe is filled alternatively with a k-essence fluid generated by 
the linear kinetic function or with quintessence.
Here, we have extended these investigations to the anisotropic BTI space time. 
We have solved the Einstein equations and found their  general solution when the k field 
is driven by an inverse square potential. We have shown that the general solution behaves asymptotically as a 
power law, allowing us to use the model to describe an average accelerated expansion. 
In addition, we have proved the kinematical equivalence of this anisotropic cosmology 
with the BTI quintessence model driven by an exponential potential in a similar way as 
it was done in a FRW universe.



\section*{Acknowledgement}  

L.P.C. is partially funded by the University of Buenos Aires  under
project X224, and the Consejo Nacional de Investigaciones Cient\'{\i}ficas y
T\'ecnicas.

\end{document}